\documentclass[prb,aps,twocolumn,superscriptaddress,showpacs]{revtex4-1}
\usepackage[OT4]{fontenc}

\usepackage{graphicx}
\usepackage{amsmath,bm,amsfonts}

\newcommand{\rl}{\rangle\!\langle}

\DeclareMathOperator{\tr}{Tr}
\begin{document}

\author{Piotr Kaczmarkiewicz}
 \email{piotr.kaczmarkiewicz@pwr.wroc.pl}
\author{Pawe{\l} Machnikowski}
 \affiliation{Institute of Physics, Wroc{\l}aw University of
Technology, 50-370 Wroc{\l}aw, Poland}
\author{Tilmann Kuhn}
\affiliation{Institut f\"ur Festk\"orpertheorie, Westf\"alische Wilhelms-Universit\"at, 48149~M\"unster, Germany }

\title{Double quantum dot in a quantum dash: optical properties}

\begin{abstract}
We study the optical properties of highly anisotropic quantum dot structures (quantum dashes)
characterized by the presence of two trapping centers located along the 
structure. Such a system can exhibit some of the 
properties characteristic for double quantum dots. We show that 
sub- and super-radiant states can form for certain quantum dash
geometries, which is manifested by a pronounced transfer of intensity between spectral lines, accompanied by the appearance of strong electron-hole correlations.
We also compare exciton absorption
spectra and polarization properties of a system with a single and double
trapping center and show how the geometry of multiple trapping centers influences
the optical properties of the system. 
We show that for a broad range of trapping geometries
the relative absorption intensity of the ground state 
is larger than that of the lowest excited states, 
contrary to the quantum dash systems characterized by a single trapping center.
Thus, optical properties of these structures are determined by fine details of their morphology.

\end{abstract}

\pacs{63.20.Kd, 71.38.-k, 73.21.La, 78.67.Hc}

\maketitle

\section{Introduction}

Quantum dashes (QDashes) are highly elongated quantum dot structures 
characterized by high asymmetry, formed spontaneously in a process of 
self-assembled growth \cite{reithmaier07} or by means of droplet epitaxy \cite{jo10}. 
Broad gain, high degree of tunability and, in some cases, a high surface 
density \cite{reithmaier07,lelarge07,djie08,dery04,sauerwald05,frechengues99,hein09}
make their optical properties favorable over other quasi-zero dimensional structures for 
many telecommunication applications. InP QDash structures are now 
commonly used in high performance lasers and optical amplifiers operating 
at 1.55~$\mu$m \cite{reithmaier07,lelarge07,rosales11}. They also show some 
promises for possible future single-photon technologies.

Apart from this practical importance, QDashes show many interesting 
features that distinguish them from more isotropic structures (quantum dots), 
in particular when their optical properties are concerned.
Some of these optical properties have been 
studied previously for uniform \cite{miska04,wei05,musial12b} and 
non-uniform \cite{musial12,kaczmarkiewicz12} QDashes. 
It has been shown that the width and 
height variation of such a structure \cite{kaczmarkiewicz12} leads to 
a strong carrier trapping in a volume much smaller
than the volume of the whole structure.
For a QDash with a single trapping center, the additional confinement leads to
a strong energy shift of the excitonic ground state, nontrivial changes in oscillator
strengths and a strong decline of the degree of linear polarization of the exciton ground state.

The morphology of some of the InAs QDash structures \cite{reithmaier07} 
suggests that variations in their width may lead to the appearance of 
multiple trapping centers within the volume of a QDash. 
It has been shown that the optical properties of the somewhat similar system of a  double quantum dot (QD), 
differ considerably from those of a single QD.\cite{stinaff06} 
Therefore one would expect 
that the physical properties of a QDash with a double trapping center (DTQDash) are different
from the properties of a QDash with a single trapping center (STQDash).
Double QDs have been the subject of extensive studies both, within an effective approach of
coupled two-level systems \cite{bayer01,sitek09b} as well as by more detailed methods of semiconductor physics.\cite{korkusinski01,gawarecki10}
DTQDash structures however, have not been studied so far.

In this paper, we study a QDash system where the presence
of height and width fluctuations leads to the appearance of the second trapping center
and study the qualitative and quantitative differences between single and double trapping system.
Then, we characterize how different geometrical factors influence electronic
and optical properties of DTQDashes. 

We show that some of the optical properties (energy spectra and line intensities) of a DTQDash 
differ considerably from a STQDash. 
The transfer of oscillator strength between the lines, in particular from the first 
excited to the ground state, is particularly strong in a relatively broad 
range of parameters where the two trapping centers are of similar magnitude. 
This results from strong electron-hole correlations, which correspond to the formation of a superradiant\cite{scheibner07} ground state.
The effect of superradiance for double QDs has been studied before within the effective model of coupled two-level systems \cite{sitek09b} borrowed from quantum optics.
In this paper, the effect of superradiant spontaneous emission is studied on the semiconductor level, using the wave functions of the QDash nanostructure.

We find out that the polarization properties are only weakly affected by the presence of the second trapping center.
Such a relatively weak dependence of the degree of linear polarization on the exact QDash shape geometry allows
one to accurately model the ensemble polarization properties even by a simple STQDash model, although
when describing other optical properties of a QDash ensemble one might 
need to take into account realistic shapes of the QDash, as their properties strongly depend on the fine details of structure morphology.


The paper is organized as follows. In Sec. \ref{sec:model},
the general theoretical framework of our study is introduced. In Sec. \ref{sec:wyn},
we present the results of theoretical modeling. First, we address the 
general differences in optical properties between a single trapping center and 
a double trapping center QDash, then we study how geometrical properties of the trapping centers influence the energy spectra,
transition intensities and polarization properties and finally we show the formation of 
sub- and super-radiant states in the system. We conclude the paper in Sec. \ref{sec:conclusion}.

\section{The model}\label{sec:model}

\begin{figure}[tb]
\begin{center}
\includegraphics[width=85mm]{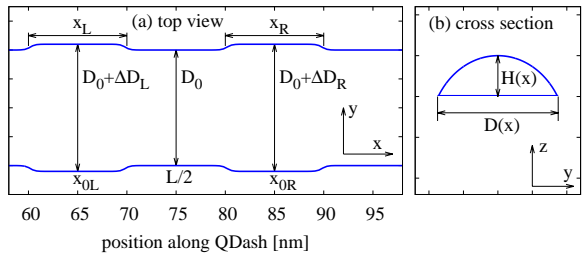}
\end{center}
\caption{\label{fig:schem} The schematic shape of QDash with two
width fluctuations present. The cross section (b) of a QDash is a
circular segment with constant height to width ratio.}
\end{figure}

We consider a highly elongated quantum dot-like structure characterized 
by the width and thickness variation at certain positions along its 
length. The width of the QDash base changes according to
\begin{equation}\label{wzor:szer}
D(x) = D_{0}+\Delta_{\mathrm{L}}(x)+\Delta_{\mathrm{R}}(x),
\end{equation}
with
\begin{displaymath}
\Delta_{\mathrm{L(R)}}(x)=\frac{\Delta D_{\mathrm{L(R)}} (1+4e^{-b})}{1+4e^{-b}\cosh[2b(x-x_{0\mathrm{L(R)}})/x_{\mathrm{L(R)}}]},
\end{displaymath}
where $x$ is the coordinate along the QDash structure, $D_0$ is the QDash 
base width away from the widening, $\Delta D_{L(R)}$ is the magnitude of the 
left (right) width fluctuation, $x_\mathrm{L(R)}$ is the length of the left (right) 
fluctuation and $x_{\mathrm{0L(R)}}$ is the position of the center of the left (right) widening (see also Fig. \ref{fig:schem}). 
The $b$ parameter defines the shape of the widening (we choose $b=20$). 
We define the widening parameter as the ratio of the excess width to the 
QDash width away from the trapping center, $\lambda_{L(R)}=\Delta D_{L(R)}/D_0$.  
The QDash width to height ratio is kept constant, 
$D(x) = \alpha H(x)$, with $\alpha=5.5$, which is typical for these structures
\cite{sauerwald05}. The total length of the structure is set to $150$ nm 
and the length to width ratio is $L/D = 6$. The geometry of the width fluctuations of a QDash
is shown in Fig.~\ref{fig:schem}. The shape parameters used in our calculations
are $\lambda_{L(R)}= 0.10$, $x_\mathrm{0L}=65$~nm, $x_\mathrm{0R}=85$~nm, $x_\mathrm{L(R)}=10$~nm, unless stated otherwise.
The STQDash shape parameters are identical to DTQDash with the exception of the position of the trapping center $x_0=L/2$.
 
In our approach to modeling the exciton states in such a highly elongated QDash,
we use the envelope function formalism and generalize our previous approach \cite{musial12} based on the configuration-interaction 
scheme for exciton states and perturbative treatment of hole subband mixing.
The single-band effective mass envelope Hamiltonian for a single carrier is
\begin{displaymath}
H_{c}=-\frac{\hbar^2}{2m^*_{c}} \Delta + V(\bm{r}),
\end{displaymath}
where $c$ denotes the carrier type (electron or hole), and $m^*_c$ is 
the effective mass of the carrier in a single band approximation. 
The QDash confinement potential is modeled as a three dimensional 
potential well, described by the $V(\bm{r})$ term, reflecting the shape 
of the structure [Eq.~(\ref{wzor:szer})] and the band edge offset 
between the QDash and host materials. The effective band offsets used
in calculations include the strain effects and are taken as 250~meV and 400~meV 
for electrons and holes, respectively.
In order to model single carrier envelope wave functions, we use a 
variational method and follow the adiabatic approximation \cite{wojs96}, 
as the confinement along the $x$ direction is much weaker than the 
confinement in the other directions. 

Our modeling follows the approach developed in Refs. \onlinecite{musial12,kaczmarkiewicz12}. 
In summary, we first variationally minimize the single-particle Hamiltonian,
at grid points along the $x$ direction,
\begin{displaymath}
H_{yz}= -\frac{\hbar^2}{2m^*}
\left( \frac{\partial^{2}}{\partial y^{2}}
  +\frac{\partial^{2}}{\partial z^{2}} \right)  + V(\bm{r}),
\end{displaymath}
in the class of two-dimensional harmonic oscillator ground state wave functions
\begin{eqnarray}
\lefteqn{\phi_{0}(y,z;x)=}\nonumber \\
&&\frac{1}{\sqrt{l_z(x) l_y(x) \pi }} \exp
\left\{ -\frac{[z-z_0(x)]^2}{2l^{2}_z(x)}
-\frac{y^2}{2l^{2}_y(x)} \right\}.\nonumber
\label{wzor:psiZ}
\end{eqnarray}
This variational minimization procedure allows us to obtain the set of 
variational parameters $l_y(x)$, $l_z(x)$ and $z_0(x)$ corresponding to 
the characteristic confinement lengths for the $y$ and $z$ directions, 
and to the center of the wave function along the $z$ direction, respectively. 
Then, a set of effective potentials for the direction parallel to the QDash elongation 
is generated,
\begin{equation}\label{wzor:EffPot}
\epsilon_n(x)=  \int dz \int dy
\phi_{n}^{*}(y,z;x)H_{yz}\phi_{n}(y,z;x),\nonumber
\end{equation}
where $\phi_n$ is the wave function of a 2D harmonic oscillator representing the $n$-th state 
along the $y$ direction. The previously obtained variational parameters are the same for
all $\phi_n$ states.
Next, we use the obtained effective potentials in a set of one dimensional 
eigenvalue equations describing the system state in the direction of the 
structure elongation
\begin{equation}\label{wzor:RnieEfektywn}
\left [ -\frac{\hbar^{2}}{2m^*}\frac{\partial^{2}}{\partial x^{2}}
+ \epsilon_n(x) \right ] f_{nm}(x) = E_{nm} f_{nm}(x).
\end{equation}
The complete approximate envelope wave functions are then $\psi_i(x,y,z)=\phi_{n}(y,z;x)f_{nm}(x)$, with $i$ denoting the set of quantum numbers $n$ and $m$.

We construct the excitonic product basis using the single carrier envelope wave functions.
The Hamiltonian describing a single exciton confined in the structure is
\begin{eqnarray}
H &=& \sum_{i} E_{i}^{(\mathrm{e})} |i_\mathrm{e} \rl i_\mathrm{e}| +\sum_{i}
E_{i}^{(\mathrm{h})} |i_\mathrm{h} \rl i_\mathrm{h}| \nonumber \\
  & &+ \sum_{ijkl} V_{ijkl} |i_\mathrm{e} j_\mathrm{h}\rl k_\mathrm{e} l_\mathrm{h}|,
\label{wzor:ham_ex}
\end{eqnarray}
where $E_i^{(\mathrm{e,h})}$ are the eigenenergies calculated from 
Eq. (\ref{wzor:RnieEfektywn}) and $V_{ijkl}$ are the the electron-hole interaction 
matrix elements.

In the calculations of the dipole moments, we assume that the hole states have 
mainly heavy hole character with only a small admixture from the light hole states. 
The interband dipole moment components corresponding to the transition from the ground state to the exciton 
state $\beta$ for the polarization parallel ($l$) and transverse ($t$)
to the direction of the elongation of the structure are
\begin{equation}\label{wzor:dl}
d_{l(t)}^{(\beta)} = \mp d_0 \frac{i\pm1}{2}\alpha_{\mathrm{hh}}^{(\beta)}
+ d_0 \frac{1\mp i}{2\sqrt{3}}\alpha_{\mathrm{lh}}^{(\beta)}\nonumber,
\end{equation}
where the upper and lower signs correspond to $l$ and $t$, respectively. 
The electron-hole overlap integrals for heavy ($\alpha_{\mathrm{hh}}^{(\beta)}$) and light hole ($\alpha_{\mathrm{lh}}^{(\beta)}$) states are defined as
\begin{equation}\label{wzor:defalpha}
\alpha_{\mathrm{hh}}^{(\beta)} =\sum_{ij} c_{ij}^{(\beta)}
\int d^3r \psi_{i}^{(\mathrm{h})}({\bm{r}})\psi_{j}^{(\mathrm{e})}({\bm{r}})\nonumber
\end{equation}
and
\begin{equation}\label{wzor:defalpha2}
\alpha_{\mathrm{lh}}^{(\beta)} =- \frac{1}{\Delta E_{\mathrm{lh}}} \sum_{ij} c_{ij}^{(\beta)}
\int d^3r \psi_{i}^{(\mathrm{h})}({\bm{r}})R_k \psi_{j}^{(\mathrm{e})}({\bm{r}}),\nonumber
\end{equation}
where $c_{ij}^{(\beta)}$ are coefficients for the expansion of the 
exciton state $\beta$ into the product basis states 
obtained by a diagonalization of the Hamiltonian given in Eq. (\ref{wzor:ham_ex}),
$R_k$ is the kinetic part of the Kane Hamiltonian matrix element 
coupling the spin $3/2$ heavy-hole subband with the $-1/2$ light hole 
subband \cite{andrzejewski10} and $\Delta E_{\mathrm{lh}}$ defines the 
average separation between the light and heavy hole states.

The intrinsic optical intensity of the line $\beta$ is then proportional to $|d_{l}^{(\beta)}|^2+|d_{t}^{(\beta)}|^2$
and the degree of linear polarization of luminescence is
\begin{equation}\label{wzor:dop}
DOP=\frac{|d_{l}^{(\beta)}|^2-|d_{t}^{(\beta)}|^2}{|d_{l}^{(\beta)}|^2+|d_{t}^{(\beta)}|^2},
\end{equation}
where $|d_{l(t)}^{(\beta)}|^2$ is the intensity of light polarized parallel (transverse) to the structure elongation.

The degree of quantum correlation between the electron and the hole is quantified 
in terms of the purity of the reduced state of one of the carriers, (say, the electron),
$P=\tr{\rho^2_{\mathrm{e}}}$, or the linear entropy $S_L=1-P$, where $\rho_e$ is the reduced density matrix of the electron, defined as
\begin{equation}\label{wzor:rho}
{\rho_{\mathrm{e}}}_{ij} = \sum_k c_{ik}^{(\beta)*} c_{jk}^{(\beta)}.\nonumber
\end{equation}
For a product (uncorrelated) electron-hole wave function one has $S_L=0$, 
while in a maximally entangled state $S_L$ drops down to $1/N$, 
where N is the number of available single-particle states.

\section{Results}\label{sec:wyn}

\begin{figure}[t]
\begin{center}
\includegraphics[width=85mm]{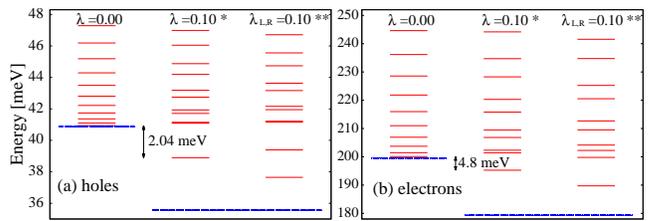}
\end{center}
\caption{\label{fig:energ} The single carrier QDash energy spectra for holes (a) and electrons (b). 
The dashed line denotes the bottom of the effective potential. 
The presented spectra correspond to a QDash with no trapping center, a STQDash with central trapping center (*) and DTQDash (**) with two identical widenings ($\Delta D_L=\Delta D_R, \lambda_L=\lambda_R$).
For the case of the DTQDash the two lowest energy states are trapped within the potential fluctuations.  }
\end{figure}


Introducing a single central width fluctuation in a QDash leads to 
the trapping of carrier ground state and strongly shifts the energies 
of states characterized by even wave function as shown in Fig.~\ref{fig:energ}.
Including the second trapping center results in an even stronger spectrum reconstruction
and leads to the trapping of the first excited state as well. 
In the case of single hole states [Fig. \ref{fig:energ}(a)],
the energy shifts of higher excited states are relatively small, though still noticeable.
Stronger energy reconstruction is visible for the lighter carrier (electron).
Not only the two lowest energy states are now trapped, but also stronger
energy shifts of higher excited states are visible [Fig. \ref{fig:energ}(b)].
Introducing the second trapping center modifies also the probability density
for finding a carrier in certain regions of the QDash. In the case of a single trapping center,
the carrier ground state is localized in the vicinity of the width fluctuation
and the first excited state occupies nearly the whole volume of the QDash, with very small
probability density in the widening area. 
In the case of a QDash with double trapping center, the ground state, as 
well the first excited state, are localized mainly in both trapping centers,
whereas the second excited state is localized outside of the trapping
centers and has a similar character as the first excited state in a STQDash.

\begin{figure}[tb]
\begin{center}
\includegraphics[width=75mm]{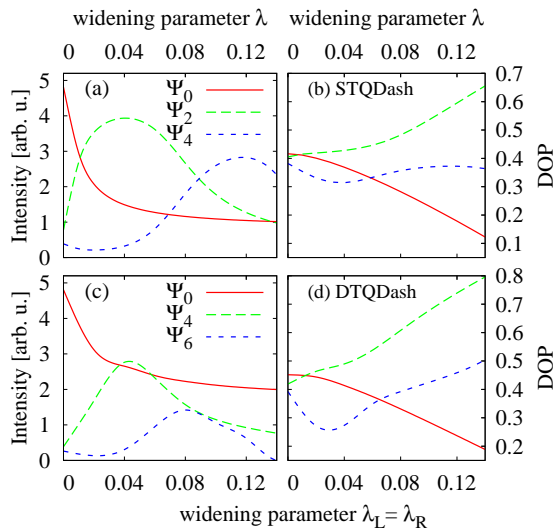}
\end{center}
\caption{\label{fig:comp} The absorption intensities (a,c) as well as the polarization
properties (b,d) for several lowest energy exciton eigenstates with large enough
transition probability as a function of trapping depth for a QDash with a single (a,b) and double (c,d) trapping.}
\end{figure}

\begin{figure}[tb]
\begin{center}
\includegraphics[width=85mm]{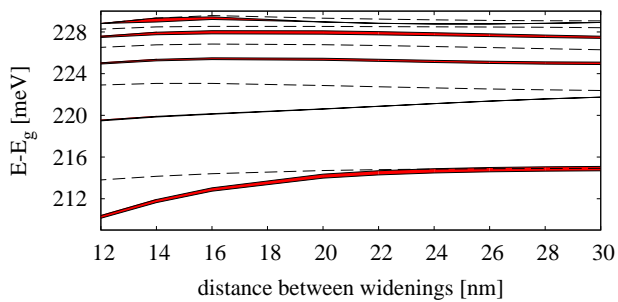}
\end{center}
\caption{\label{fig:dist-sym} The exciton energy spectrum as a function of the distance between the centers of the widenings ($D_2$ symmetry preserved).
The linewidth of the bright states is proportional to the absorption intensity. Dark states are denoted by a dashed line.}
\end{figure}
\begin{figure}[tb]
\begin{center}
\includegraphics[width=85mm]{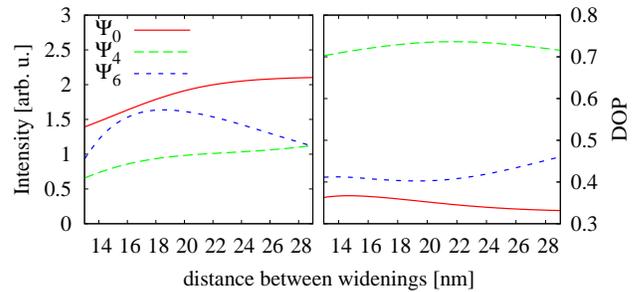}
\end{center}
\caption{\label{fig:dist-sym-dop} The absorption intensities and the degree of linear polarization as a function of the distance between the centers of two identical widenings.}
\end{figure}
In Fig.~\ref{fig:comp}, we present the absorption intensities and polarization properties of STQDash [Fig.~\ref{fig:comp}(a-b)] and DTQDash [Fig.~\ref{fig:comp}(c-d)] structures
as a function of the widening parameter $\lambda$. In Fig.~\ref{fig:comp}(c-d) two identical widenings are assumed ($\lambda_L=\lambda_R$).
We only show states with even indices, as only these states have nonvanishing electron hole overlaps.
Furthermore, in Fig.~\ref{fig:comp}(c-d), we omit the state $\Psi_2$ as it is constructed mainly of the electron ground
state and the second excited hole state. Since the ground state of the electron occupies the area
of the QDash widenings and the second excited hole state the area away from the widenings their
overlap is also very small and this state does not contribute significantly to the optical spectra of the system.
One can see that the qualitative features of the three presented optically active states are very similar.
On the other hand, one can also notice that there is a rather significant change in the line intensities.
An enhancement of the absorption intensity of the exciton ground state in the DTQDash compared to the STQDash for moderate and large values
of the widening parameter is visible, accompanied by a decline in the intensities of the excited states.
The intensities of the excited states are no longer stronger than that of the ground 
state, apart from one particular range of widening parameter (around $\lambda = 0.04$).
The decline in the intensity of the excited states is followed by the enhancement of the intensity of the ground state.
This enhancement is present even for small values of the widening ($\lambda > 0.02$), and is most pronounced for 
larger widenings, for which the intensity of the ground state for the DTQDash is up to two times larger than in the STQDash case.
The polarization properties are very similar for both the STQDash and DTQDash [Fig. \ref{fig:comp}(c-d)].
Qualitatively there is no difference between those two, only a small shift of about 0.1 is
observed for large values of the widening parameter. Since the changes in the polarization properties of the structure are so small, 
one can conclude that the presence of the second width fluctuation will 
have limited impact also on the ensemble polarization properties.

So far, we have analyzed the new features appearing in the QDash absorption spectra assuming identical and symmetrically located trapping centers. However, 
in order to fully characterize the optical properties one has to include more general variations of the 
parameters defining the geometry of the trapping centers.

We choose three different parameters defining the trapping centers within a QDash
and present the energy spectra, line intensities, as well as the polarization properties of the system.
We study the effect of changing the following parameters: 
the distance between the trapping centers,
the depth ($\lambda_R$), and the length ($x_{R}$) of one of the trapping centers.
In the first case the $D_2$ symmetry of the structure is preserved while in the other two cases it is broken.
The study of changing the distance between the trapping centers corresponds to changing the tunnel coupling between two QDs,
while the two latter cases correspond to driving the energies of the two QDs formed within a QDash through the resonance.
\begin{figure}[tb]
\begin{center}
\includegraphics[width=85mm]{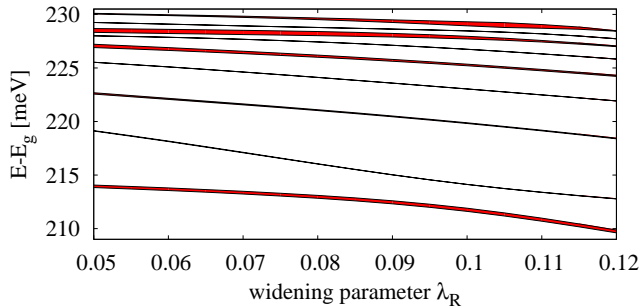}
\end{center}
\caption{\label{fig:ssup1} The exciton energy spectrum as a function of the value of the widening parameter $\lambda_R$ with constant $\lambda_L$ = 0.10. 
An anti-crossing feature can be observed at $\lambda_R$ = 0.10. The linewidth if proportional to the absorption intensity. The positions of widenings are $x_L=68$~nm and $x_R=82$~nm.  }
\end{figure}
\begin{figure}[tb]
\begin{center}
\includegraphics[width=85mm]{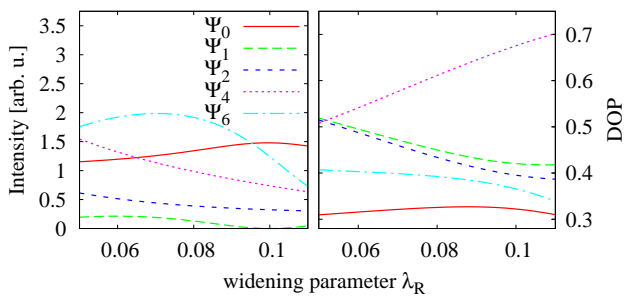}
\end{center}
\caption{\label{fig:ssup1-dop} The absorption intensities (a) and the degree of linear polarization (b) as a function of the value of the widening parameter $\lambda_R$.
For $\lambda_{R}$ = 0.10 a formation of sub- ($\Psi_1$) and super-radiant state ($\Psi_0$) can be observed. }
\end{figure}

In Fig. \ref{fig:dist-sym}, we show the energy spectrum of an exciton confined in a QDash as a function of the separation between identical widenings.
As can be observed, the gap between the two lowest energy exciton states closes when the separation between trapping centers increases.
For small values of the distance, the absorption intensity of the second excited state slightly increases.
This is due to the fact that, for small enough separations, the carrier probability density does not vanish completely in between the trapping centers, and envelope wave functions are close to the ones
for the case of a STQDash. The separation between the widenings has relatively small impact on the intensities of the brightest low energy exciton states [Fig. \ref{fig:dist-sym-dop}(a)] with changes smaller
than by a factor of 2. Also, the changes in DOP are rather insignificant [Fig. \ref{fig:dist-sym-dop}(b)], with DOP of consecutive states being low ($<0.4$), high ($>0.7$) and low again ($<0.5$), respectively.
Such a relation between the subsequent exciton states leads to a characteristic $S$-shaped temperature dependence of the DOP\cite{kaczmarkiewicz12}.
\begin{figure}[tb]
\begin{center}
\includegraphics[width=85mm]{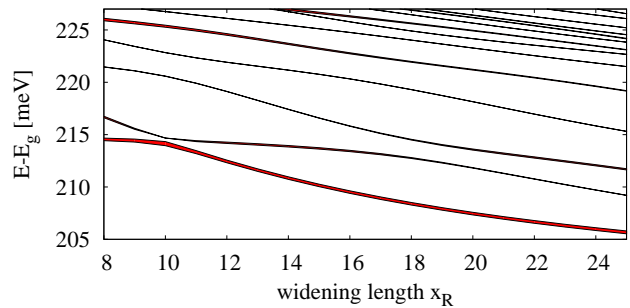}
\end{center}
\caption{\label{fig:ssup2} The exciton energy spectrum as a function of the right widening length $x_R$. The left widening length is set to $x_L$~=~10~nm.
The linewidth if proportional to the absorption intensity.}
\end{figure}
\begin{figure}[tb]
\begin{center}
\includegraphics[width=85mm]{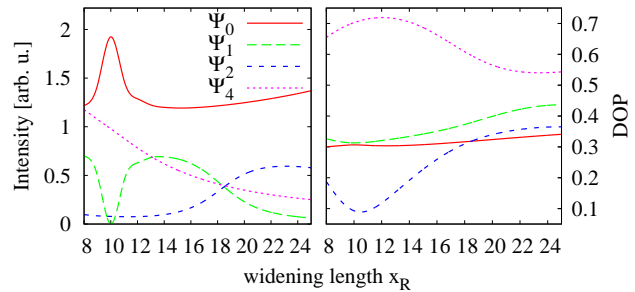}
\end{center}
\caption{\label{fig:ssup2-dop} The absorption intensities for several lowest energy exciton eigenstates (a) 
and the degree of linear polarization (b) as a function of the widening length $x_R$. A strong enhancement
of the ground state ($\Psi_0$) and decline in the intensity of the first excited state ($\Psi_1$) can be observed around the point of $\lambda_L=\lambda_R$.}
\end{figure}

In Fig. \ref{fig:ssup1}, we show the energy spectrum as a function of 
the right widening amplitude $\lambda_R$. 
One can see a decline in the energies of the exciton eigenenergies for
larger values of the widening amplitude, and an anticrossing-like
feature can be observed close to $\lambda_R$ = 0.10. A similar anticrossing would be observed
in the case of two interacting QDots with similar sizes.
A characteristic feature of such a system is a decline in the intensity of
one of the states [$\Psi_1$ in Fig.~\ref{fig:ssup1-dop}(a)]
and enhancement of the intensity of the other state [$\Psi_0$ in Fig.~\ref{fig:ssup1-dop}(a)]. 
In the two-QDot system the bright
state is $(|1\rangle+|2\rangle)/ \sqrt{2}$, where $|n\rangle$ denotes the exciton
occupying the $n$-th QDot, and the corresponding dark state is $(|1\rangle-|2\rangle)/ \sqrt{2}$.
Here, for the resonance condition, the state $\Psi_0$ is a superposition of the basis states
with electron and hole part of the same parity which has non-zero electron-hole overlap (bright state), while the state $\Psi_1$ 
is a superposition of basis states with electron and hole wave function of opposite parity and is therefore a dark state.
The observed polarization properties [Fig. \ref{fig:ssup1-dop}(b)] of the exciton ground state ($\Psi_0$) change very little with $\lambda_R$. Higher exciton
states exhibit a stronger change in the DOP, but the general picture where the ground state has a relatively small DOP and the DOP of
higher energy states is large still holds true.
The energy splitting for the resonance condition is here 2.4 meV and the effective coupling $t_{\mathrm{eff}}=1.2$~meV is
greater than the value calculated for vertical QDot molecule systems of $0.7$~meV \cite{gawarecki12}.
This is a result of a lower potential barrier between the trapping centers within the QDash compared
to the potential barrier between stacked quantum dots.

Similar features are observed when, instead of the widening parameter, the widening length ($x_R$) is varied.
As can be seen in Fig.~\ref{fig:ssup2}, the energy separation between the excitonic states in this case changes in a non-trivial way. 
Also here, a decline in the energy of the presented states can be observed for increasing $x_R$, which is a result
of decreasing the average effective potential. One can notice that an anticrossing for values of $x_R$ close to 10~nm is visible in the spectrum. 
For such a value of this parameter, the shape of the left and right widenings are identical and a characteristic 
enhancement of the intensity of the ground state and decline in the intensity of the first excited state 
is clearly visible [Fig.~\ref{fig:ssup2-dop}(a)]. Here, the effect is much more sensitive to the widening length than to the widening amplitude,
although one has to keep in mind that this depends also on the distance between the widenings, which was not the same.
Again, such a behavior results from the formation of a sub- and superradiant state.

The change in the DOP shown in Fig.~\ref{fig:ssup2-dop}(b) is nonmonotonic, but again the DOP of the ground state is nearly constant
in a very broad range of $x_R$. It is also worth noticing that the DOP of the second
excited state ($\Psi_2$) can be much lower than that of the ground state (for small widening lengths),
but the intensity of this state  is very low for these values of $x_\mathrm{R}$ and it is not
expected to have significant impact on overall optical properties of QDash systems.

\begin{figure}[tb]
\begin{center}
\includegraphics[width=85mm]{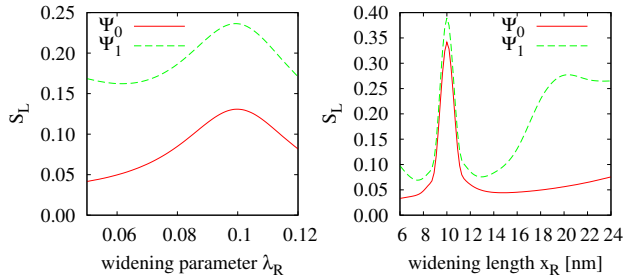}
\end{center}
\caption{\label{fig:korelacje} Linear entropy of the ground state and the first excited state as a function of widening parameter $\lambda_R$ (a) and widening length $x_R$ (b).}
\end{figure}
 
While an electron-hole pair confined in a single quantum dot can be approximated by a product wave function, 
in a double-dot structure near resonance the excitation is coherently delocalized as a whole between the two dots, 
which means that the two carriers become correlated (entangled). In order to study the degree of this correlation,
in Fig.~\ref{fig:korelacje}, we show the linear entropy for the ground and 
the first excited state as a function of the widening parameter and widening length.
When super- and sub-radiant states form, a strong increase in electron-hole correlation is visible.
This correlation effect is even much more pronounced than the intensity transfer seen in the intensity curves in Fig.~\ref{fig:ssup1-dop}(a).
In Fig.~\ref{fig:korelacje}(b) for a widening length $x_R=10$~nm the impurity of the ground state reaches 0.34
which translates to a strongly mixed state (the limiting value for an effective two-level system corresponding to the two trapping sites is 0.5).
The linear entropy of the first excited state is higher than of the ground state, as it
interacts more strongly with higher energy states, which also results in stronger correlations.

\section{Conclusions}\label{sec:conclusion}
In this paper, we have investigated the influence of the presence of two trapping centers
within a QDash on its optical and polarization properties. We have shown
that even though the presence of the second trapping center strongly
changes the spectrum of the system, the general polarization properties
are similar to the properties of a QDash with a single trapping center.
The relative absorption intensities of the lowest excited states for DTQDash are reduced compared
to the STQDash case. For a DTQDash system, the intensity of the ground state is either significantly larger or comparable
than that of the excited states.
We have investigated the effect of symmetry breaking of a QDash, by changing
the widening length and amplitude. Both these parameters have significant influence
on the electronic and optical properties of the system. Depending on the exact values of
the shape parameters of a QDash widening one can observe strong changes in both absorption intensities
and polarization properties. For values of QDash shape parameters which correspond to two similar trapping centers, 
enhancement of the intensity of the ground state at the expense of the intensity of the first excited state has been observed.
This effect is analogous to the formation of super- and sub-radiant states in a system of two quantum emitters, which 
appears as a result of reconstruction of the wave functions and building strong electron-hole correlations.
Also, the energy splitting between the ground and excited state for the system with two identical trapping centers
is larger than that observed for stacked QD molecules. This is due to the
fact that the potential barrier between the trapping centers in a DTQDash is lower
than that between two stacked quantum dots.
Our results show that some properties of the optical emission from a QDash (energy spectra, relative line intensities)
are strongly affected by its detailed morphology features, hence in order
to fully characterize the optical properties of a QDash ensemble the information
about QDash shape distribution might be necessary. On the other hand, the polarization properties of emitted light
are almost insensitive to the presence of multiple trapping centers, therefore the theory based on a simple STQDash model\cite{musial12},
describing the degree of linear polarization observed in emission from ensemble of QDashes, remains valid also for more realistic structures.

\section{Acknowledgments}
We acknowledge support from the TEAM programme of the 
Foundation for Polish Science, co-financed by the European Regional
Development Fund, and from the Alexander von Humboldt Foundation.
PK acknowledges support from the German Academic Exchange Service (DAAD).
Calculations have been partly carried out in Wroclaw Centre for Networking and Supercomputing
(http://www.wcss.wroc.pl), grant No. 203.

\bibliographystyle{prsty}
\bibliography{abbr,quantum}

\end{document}